\documentclass[runningheads]{llncs}
\usepackage[T1]{fontenc}
\usepackage{graphicx,verbatim}
\usepackage{amsmath}
\usepackage{hyperref}
\usepackage{subcaption}

\begin{document}
\newcommand{\method}{\texttt{DisMorph}}
\title{{\method}: learning to disentangle technical distortions from true biological change}
\titlerunning{{\method}: disentangling technical distortions from biological change}
%
%
\author{
Jingru Fu\inst{1-3,}\thanks{Corresponding author.} \and 
Kathleen E.~Larson\inst{1-3} \and 
Douglas N.~Greve\inst{1-3} \and 
\\
Bruce Fischl\inst{1-3}\textsuperscript{,\ensuremath{\dagger}} \and 
Malte Hoffmann\inst{1-3}\textsuperscript{,\ensuremath{\dagger}} 
}
%
\authorrunning{J. Fu et al.}
%
\institute{
Athinoula A. Martinos Center for Biomedical Imaging,
Charlestown, USA 
\and
Department of Radiology, Massachusetts General Hospital,
Boston, USA 
\and
Department of Radiology, Harvard Medical School,
Boston, USA\\ 
\email{jifu1@mgh.harvard.edu} 
}
\maketitle              
\begingroup
\renewcommand{\thefootnote}{\ensuremath{\dagger}}
\footnotetext{Shared senior authorship.}
\endgroup
\begin{abstract}
Longitudinal MRI enables sensitive measurement of structural brain change for studying aging and neurodegenerative disease. Deformable image registration is a key tool for estimating such change by computing a dense deformation that captures geometric differences between longitudinal scans. However, MRI scanners introduce geometric distortions that vary across acquisition systems and protocols, such as gradient non-linearity (GNL) distortion.
Existing registration methods estimate a single field that conflates biological and technical effects, potentially biasing downstream morphometric measurements if distortions remain (partially) uncorrected.
We propose \method, a registration framework trained entirely on synthetic data that explicitly decomposes longitudinal deformation into \textit{technical} and \textit{anatomical} transforms. \method{} predicts two dense deformations, each encoding one effect. During training, a novel generative model synthesizes both effects separately to provide disentanglement supervision, while domain randomization promotes generalization across imaging protocols.
We evaluate \method{} in three complementary settings. On simulated data with known ground truth, \method{} detects anatomical change more accurately and consistently than conventional registration. On real image pairs that differ only by GNL distortion, \method{} assigns most geometric change to the distortion field, demonstrating specificity in the absence of anatomical change. On longitudinal Alzheimer's disease (AD) pairs, \method{} detects anatomical change in AD-related brain structures while identifying residual distortion left after standard correction.
By disentangling MRI-induced distortion from biological change in the longitudinal deformation, \method{} paves the way for more accurate longitudinal morphometry in clinical settings where maintaining acquisition consistency is challenging.

\keywords{Longitudinal morphometry \and Deformable registration \and Neuroimaging \and Domain randomization \and Gradient non-linearity distortion.}
\end{abstract}
\section{Introduction}

\begin{figure}[t]
\centering
\includegraphics[width=0.8\textwidth]{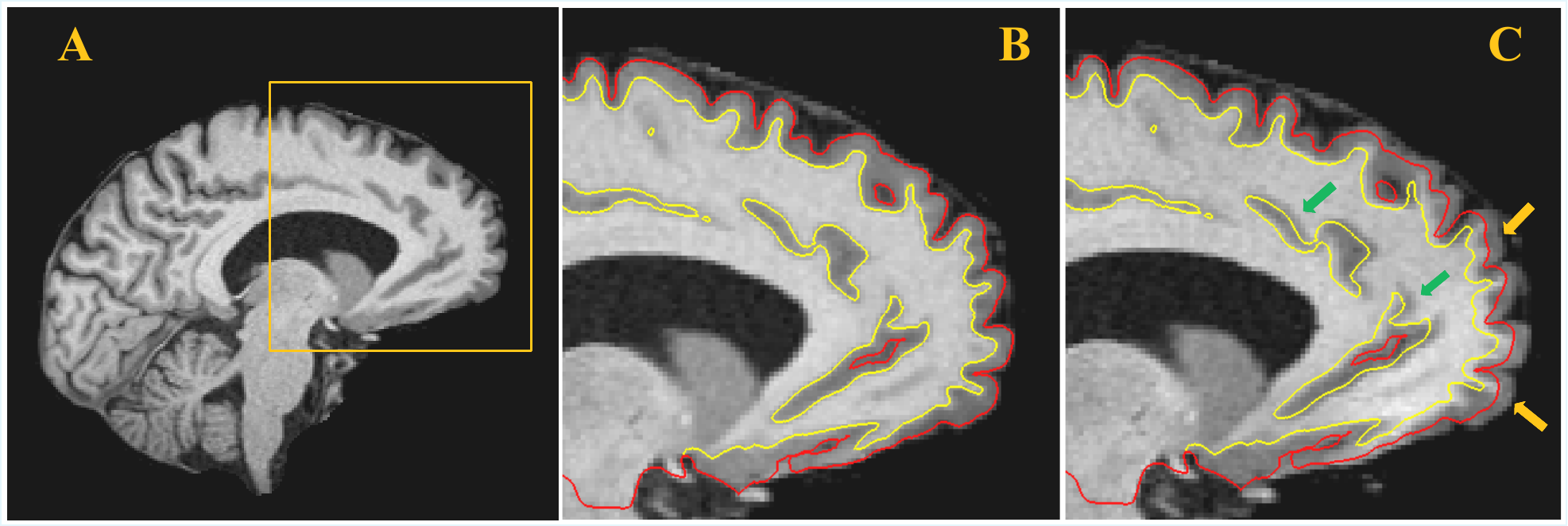}
\caption{Tissue boundary shifts induced by gradient non-linearity in MRI. \textbf{(A)}~Sagittal fBIRN source slice~\cite{keator_function_2016}. 
\textbf{(B)} Pial (red) and white-matter (yellow) surfaces fitted to the distortion-corrected image. 
\textbf{(C)} The same surface overlays on the original, \textit{distorted} image. 
Arrows highlight boundary displacements.
} \label{fig:distortion}
\end{figure}

Morphometric analysis of brain magnetic resonance imaging (MRI) enables in-vivo quantification of structural neuroanatomical properties indicative of neurodevelopment, neurodegenerative disease, and aging trajectories~\cite{ashburner2000voxel,good2001voxel,fu2025decomposing}. Compared to cross-sectional imaging, longitudinal imaging improves sensitivity by eliminating inter-subject anatomical variability and enables within-subject comparisons over time~\cite{risacher2010longitudinal,reuter_highly_2010,reuter_within-subject_2012}. Accurate estimation of longitudinal change is therefore fundamental to brain morphometry and disease progression studies.
Unfortunately, imperfections in MRI hardware introduce undesirable, non-anatomical distortions. Differences between imaging systems and acquisition protocols cause these distortions to vary across longitudinal scans, which can obscure true biological change, thereby reducing statistical power and biasing estimates of disease progression~\cite{jovicich2006reliability,van_der_kouwe_brain_2008,takao2010computational}. Spatial distortions are particularly problematic for morphometry because they directly affect measurements of volume~\cite{fischl2002whole}, shape~\cite{miller2004computational}, and tissue boundaries~\cite{barnes2004differentiating}.

Gradient non-linearity (GNL) is the most prominent source of distortion in structural MRI (Fig.~\ref{fig:distortion})~\cite{jovicich2006reliability}.
Although GNL correction is available on modern MRI systems, this correction is not always applied and can be incomplete: residual distortion can persist even after vendor correction, with reported magnitudes of up to 2 mm within a typical brain-imaging field of view, comparable in scale to annual atrophy rates measured in neurodegeneration studies~\cite{alzahrani2020audit,nousiainen2020measuring}. Furthermore, retrospective datasets often provide no reliable way of verifying whether studies applied GNL correction correctly or at all. For instance, ADNI phantom monitoring identified issues including GNL correction with coefficients for the wrong gradient system, disabled autoshimming, and a miscalibrated laser alignment light, which would have compromised morphometry at over 25\% of participating sites had they remained undetected~\cite{gunter2009measurement}. As an alternative to vendor correction, phantom-based calibration compares image geometry to a known phantom geometry. Although useful, it is impractical for retrospective, legacy, or large multi-site studies. Therefore, image-based approaches capable of separating residual MRI-induced distortion from biological change in longitudinal image pairs would add value and complement vendor-provided GNL correction.

A core technique for quantifying longitudinal change is deformable image registration, which estimates a dense deformation field that captures geometric differences between scans.
Classical registration has undergone decades of development~\cite{avants_symmetric_2008,ashburner_image_2000,thirion_image_1998,yushkevich_bias_2010}. More recently, deep-learning methods have substantially improved computational efficiency and registration accuracy~\cite{de2019deep,dalca_unsupervised_2019,balakrishnan_voxelmorph_2019,hoffmann_synthmorph_2022,gopinath2024registration,tian2024unigradicon}. Among these, several works ~\cite{hoffmann_synthmorph_2022,hoopes_synthstrip_2022,billot_synthseg_2023,iglesias_synthsr_2023,hoffmann_anatomy-aware_2024,hoffmann_domain-randomized_2025,abulnaga2025multimorph} use highly variable synthetic training images to enhance robustness to image contrasts, resolutions, and acquisition conditions. However, most learning approaches focus \textit{exclusively} on inter-subject alignment rather than longitudinal change~\cite{fu_learning_2025}. More importantly, both classical and learning-based registration methods aim to explain \textit{all} spatial differences in a single deformation, without the ability to disentangle technical distortions from true biological change.

In this study, we introduce \method{}, a learning-based registration framework that explicitly disentangles MRI distortions from biological change.
It uses a novel generative strategy to synthesize diverse spatial effects---GNL distortion and atrophy---together with intensity variations that make the registration robust to real-world variability in acquisition hardware and protocols.
We demonstrate in controlled simulations that \method{} recovers the underlying sources of three-dimensional (3D) geometric effects, and show on real datasets that its decompositions are consistent with the expected technical and anatomical differences, enabling reduction of distortion-related bias for downstream analysis.

\begin{figure}[!t]
\includegraphics[width=\textwidth]{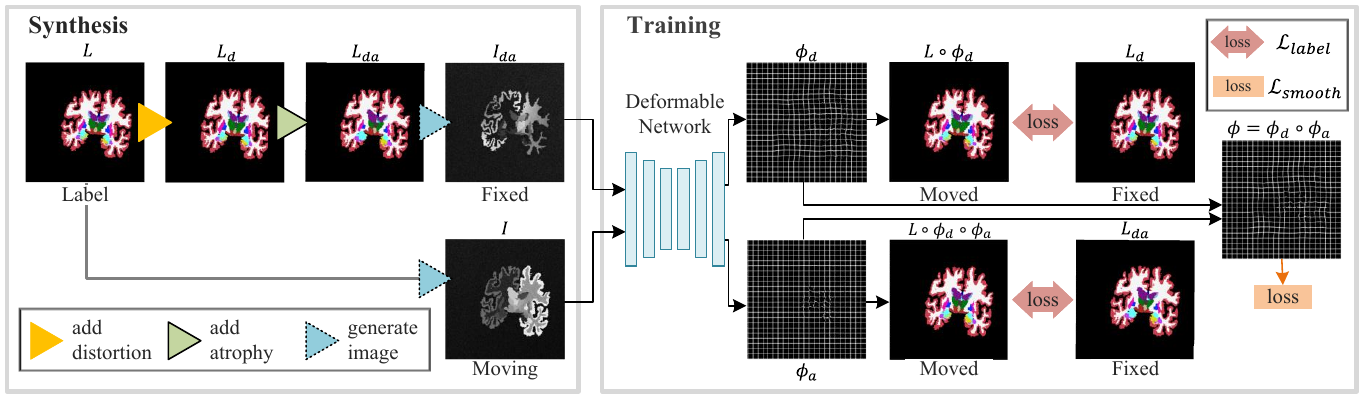}
\caption{{\method} strategy. We synthesize longitudinal training pairs with a geometry generator that separately applies distortion and atrophy to label maps, followed by an intensity generator that synthesizes images with corruptions from them. A 3D registration model learns to disentangle both geometric effects.} \label{fig:method}
\end{figure}

\section{Methods}

The proposed framework, \method{}, combines a procedural, generative model that synthesizes longitudinal image pairs for training and a registration network capable of disentangling these effects by predicting separate deformation fields for distortion and atrophy. Fig.~\ref{fig:method} provides an overview of our strategy.

\subsection{Synthesizing Longitudinal Training Data}\label{generators}
\begin{figure}[t]
\includegraphics[width=\textwidth]{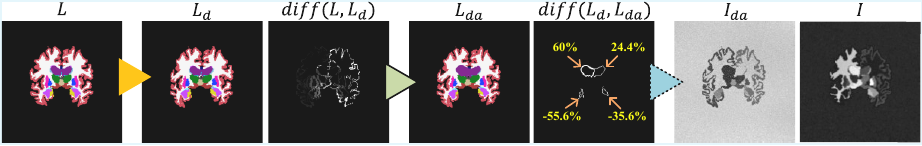}
\caption{Example of randomized synthesis steps (Fig.~\ref{fig:method}). The realization shows hippocampal atrophy and ventricular enlargement across both hemispheres.} \label{fig:synthesis}
\end{figure}

{\method} synthesizes images of diverse geometry and contrast from anatomical label maps using two procedural models: 
(i) a \emph{geometry generator} that spatially augments label maps and
(ii) an \emph{intensity generator} that creates variable gray-scale images from them.  
These models synthesize diverse deformations, image contrasts, and corruptions on the fly. Fig.~\ref{fig:synthesis} illustrates the data generation.

\noindent\textbf{Geometry generator.} Let $L$ denote a multi-class anatomical label map sampled from a training dataset $\mathcal{D} = \{L^{(n)}\}_{n=1}^{N}$, delineating $K$ distinct brain structures. We spatially augment $L$ by applying a nonlinear spatial transform.
To simulate compound longitudinal effects, we employ two additional spatial transformations.
First, we apply a smooth random elastic transform to $L$ that models GNL distortion, yielding the distorted label map $L_d$. We generate this transform by sampling displacement vectors on a coarse control lattice and interpolating it to full resolution. We sample vector magnitudes up to 2.5\% of the field of view and lattice resolutions between 2 and 10 control points for each dimension, varying the smoothness of the vector field.
We integrate the resulting field using \textit{scaling and squaring} (SS)~\cite{arsigny_log-euclidean_2006} to obtain a diffeomorphism. This stochastic generation produces diverse distortions.

Next, we introduce anatomical changes by moving label boundaries in $L_d$. Specifically, we choose a label subset $\mathcal{K} \subset \{1,\dots,K\}$ of brain regions and simulate local atrophy or growth through iterative boundary reassignment between each target structure and its neighboring labels. For each label $k \in \mathcal{K}$, we uniformly sample a volumetric change ratio $r_k \sim \mathcal{U}(-0.5,\,0.5)$, defined as $r_k = V_k^\text{target} / V_k^\text{orig} - 1$, where negative values simulate atrophy, and positive values simulate growth. We use distance maps to identify boundary voxels shared with adjacent structures and reassign these voxels to progressively shrink or expand the target region until the resulting volume change matches or exceeds $r_k$. Adding distortion and tissue change yields the final label map $L_{da}$.

\noindent\textbf{Intensity generator.} We generate intensity images $(I, I_{da})$ from label maps $(L, L_{da})$ following prior synthesis-based approaches~\cite{billot_synthseg_2023,hoffmann_anatomy-aware_2024,fu_learning_2025}. For each image $I$, briefly, we assign a different, random mean intensity $\mu_k$ to voxels $x$ associated with each label $k$ and sample intensities normally, $I(x) \sim \mathcal{N}(\mu_k, \sigma_k^2)$, where \(\mu_k \sim \mathcal{U}(0,1)\) and \(\sigma_k \sim \mathcal{U}(0,0.05)\). Next, we apply a set of image corruptions, including bias fields, gamma transforms, blurring, and additive noise~\cite{billot_synthseg_2023,fu_learning_2025}.

\subsection{Disentangling Distortion and Atrophy}\label{decompose}
We train a registration model using synthetic intensity images as input and supervise it with the corresponding label maps. This strategy removes dependence on specific image contrast and encourages the model to focus on anatomy and geometry instead. After training, the model can register images without requiring label maps. Let $I_{\textrm{F}}$ and $I_{\textrm{M}}$ denote a fixed and a moving image, respectively, with corresponding label maps $L_{\textrm{F}}$ and $L_{\textrm{M}}$. The network predicts two stationary velocity fields (SVFs), one capturing MRI distortions ($v_d$) and the other capturing anatomical change ($v_a$). To obtain diffeomorphisms, we integrate each SVF via SS, yielding deformation fields $\phi_d = \exp(v_d)$ and $\phi_a = \exp(v_a)$ that warp the moving image towards the fixed image. We represent label maps using one-hot encoding, denoted by $\tilde{L}$, and train the model by optimising the mean squared error between label representations. Given a deformation $\phi$, we define the loss as
\begin{equation}
    \mathcal{L}_\text{label}(\tilde{L}_{\textrm{M}},\tilde{L}_{\textrm{F}};\phi) =
    \frac{1}{|\Omega|} \sum_{x \in \Omega}
    \left\|
    (\tilde{L}_{\textrm{M}} \circ \phi)(x)
    -
    \tilde{L}_{\textrm{F}}(x)
    \right\|_2^2 ,
\end{equation}
where $\Omega$ denotes the spatial domain and $\tilde{L}_{\textrm{M}} \circ \phi$ is $\tilde{L}_{\textrm{M}}$ deformed by $\phi$. We use separate distortion and anatomical terms $\mathcal{L}_d=\mathcal{L}_\text{label}(\tilde{L},\tilde{L}_d;\phi_d)$ and $\mathcal{L}_a=\mathcal{L}_\text{label}(\tilde{L},\tilde{L}_{da};\phi_{da})$, respectively, and denote the composed deformation by $\phi_{da}=\phi_d\circ\phi_a$.
Let $\phi_{da}(x) = x + u(x)$ denote the displacement formulation of the deformation.
We encourage warp smoothness by adding regularization term $\mathcal{L}_{\text{smooth}}=\|\nabla u\|_2^2$, yielding the final training objective
\begin{equation}
\mathcal{L}_{\text{total}}
= \mathcal{L}_{d}
+ \mathcal{L}_{a}
+ \lambda \, \mathcal{L}_{\text{smooth}},
\end{equation}
where $\lambda$ controls the regularization strength.

\section{Experiments and Results}

We design our experiments to address two key questions. (1) To what extent can \method{} reduce errors in the estimation of known atrophy compared to conventional registration in distorted image pairs? (2) Are the two transforms predicted by \method{} sufficiently specific to disentangle GNL distortion from anatomical atrophy in real MRI data?
We address (1) in Sec.~\ref{synthexp} using simulated longitudinal MRI generated from a biomechanical atrophy model with known ground-truth tissue change. We compare the change estimated by \method{} with the known ground truth.
We address (2) in Sec.~\ref{realexp} using two complementary real MRI datasets. First, we evaluate performance on image pairs that differ only by GNL distortions. Second, we evaluate performance on longitudinal image pairs from subjects with Alzheimer's disease (AD), acquired two years apart and expected to differ primarily due to atrophy.

\noindent\textbf{Training Data.} We train {\method} on images synthesized from 530 label maps obtained by segmenting structural, T1-weighted images from IXI~\cite{ixi_dataset} and FSM~\cite{greve2024freesurfer} using FreeSurfer~\cite{fischl_freesurfer_2012,reuter_within-subject_2012,hoopes2024voxelprompt}. We affine-align all subjects to Talairach space and crop images to a fixed size of $160 \times 192 \times 224$ isotropic 1-mm voxels.

\noindent \textbf{Metrics.}
For each structure $s$, we compute the absolute symmetrized percent change (\texttt{ASPC})~\cite{reuter_within-subject_2012} between the moving and warped label maps. Specifically, we obtain sub-voxel volume estimates by interpreting interpolated one-hot values as partial-volume fractions. We compute \texttt{ASPC} as:
\begin{equation}
\mathrm{\texttt{ASPC}}_s= 
100 \times 
\frac{|V_{\text{moving}, s} - V_{\text{moved}, s}|}
{0.5 \,(V_{\text{moving}, s} + V_{\text{moved}, s})} .
\end{equation}

\noindent \textbf{Implementation Details.}
As a general registration framework, {\method} can adopt any backbone to estimate deformation fields. We choose a dual-stream encoder with shared weights and filter sizes \{32, 64, 64, 64\} to encode fixed and moving images separately. We then use a pyramid-warping encoder~\cite{jian2025disentangling} with filter sizes \{128, 128, 128, 3\} to estimate deformation fields in a coarse-to-fine manner. The network predicts the distortion field at quarter resolution and the anatomical field at half resolution. We upsample both fields to full resolution to compute the losses (Sec.~\ref{decompose}). Model training uses the Adam optimizer with batch size 1 and learning rate $10^{-4}$ for 2,600 epochs with 200 steps per epoch, after which the loss visually converges. We set the smoothness regularization weight to $\lambda=0.01$ based on empirical tuning.

\subsection{Experiments on Simulated Data} \label{synthexp}

\begin{figure}[t]
\centering
\includegraphics[width=\textwidth]{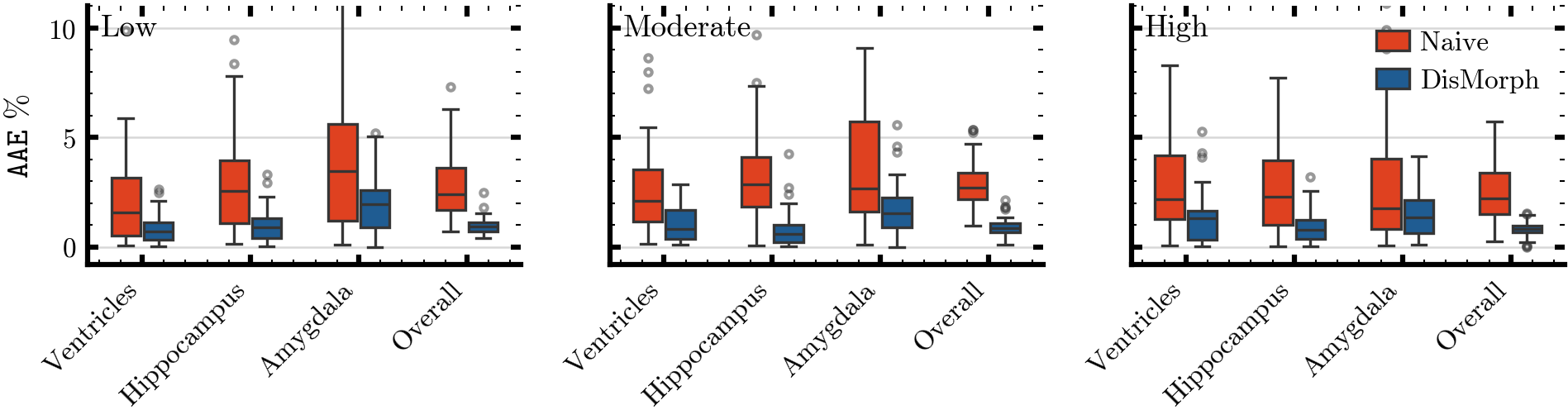}
\caption{Atrophy detection error (\texttt{AAE}) on simulated data across three atrophy levels. Lower values indicate better agreement with the ground truth.}
\label{fig:simulated_results}
\end{figure}

\noindent \textbf{Simulated Data.}
We use 150 subjects from
ADNI 1 (1.5T and 3T)~\cite{jack_jr_alzheimers_2008} and MIRIAD~\cite{malone_miriadpublic_2013}, divided equally into low-, moderate-, and high-atrophy groups. For each subject, we prescribe annual volume change rates for three AD-affected regions---ventricle,
hippocampus, and amygdala---based on prior studies~\cite{tang_diffeomorphometry_2015,fu_synthesizing_2025}.
\textbf{Low:} ventricles $(2\pm1)\%$, hippocampi $(-1\pm0.5)\%$, amygdalae $(-1\pm0.5)\%$. \textbf{Moderate:} ventricles $(3\pm1.5)\%$, hippocampi $(-1.5\pm1)\%$, amygdalae $(-1.5\pm1)\%$. \textbf{High:} ventricles $(4.5\pm2)\%$, hippocampi $(-2\pm1.5)\%$, amygdalae $(-2\pm1.5)\%$.
We simulate atrophy using a biophysical model~\cite{khanal2016biophysical}, then add distortion to create distorted longitudinal pairs.

\noindent \textbf{Setup.}
We measure atrophy estimation bias as the absolute \texttt{ASPC} error (\texttt{AAE}) between the estimated and prescribed \texttt{ASPC}. As a baseline (\texttt{Naive}), we compute \texttt{ASPC} from the
composite deformation field, which includes both anatomical change and
MRI-induced distortion. For \method{}, we instead compute \texttt{ASPC}
from the estimated atrophy field alone. This comparison measures the
benefit of disentangling MRI-induced distortion from true anatomical change.

\noindent \textbf{Results and Discussion.}
Fig.~\ref{fig:simulated_results} shows that \method{} consistently reduces \texttt{AAE} relative to \texttt{Naive} across groups, demonstrating robustness to atrophy magnitude and location. Importantly, \method{} performs best where it matters most: low atrophy corresponds to early disease stages, where distortion removal yields the largest reduction in estimation bias~\cite{risacher2010longitudinal}.

\subsection{Experiments on Real-World Data}\label{realexp}
\begin{figure}[t]
\centering
\begin{subfigure}[t]{0.495\textwidth}
    \centering
    \includegraphics[width=\linewidth]{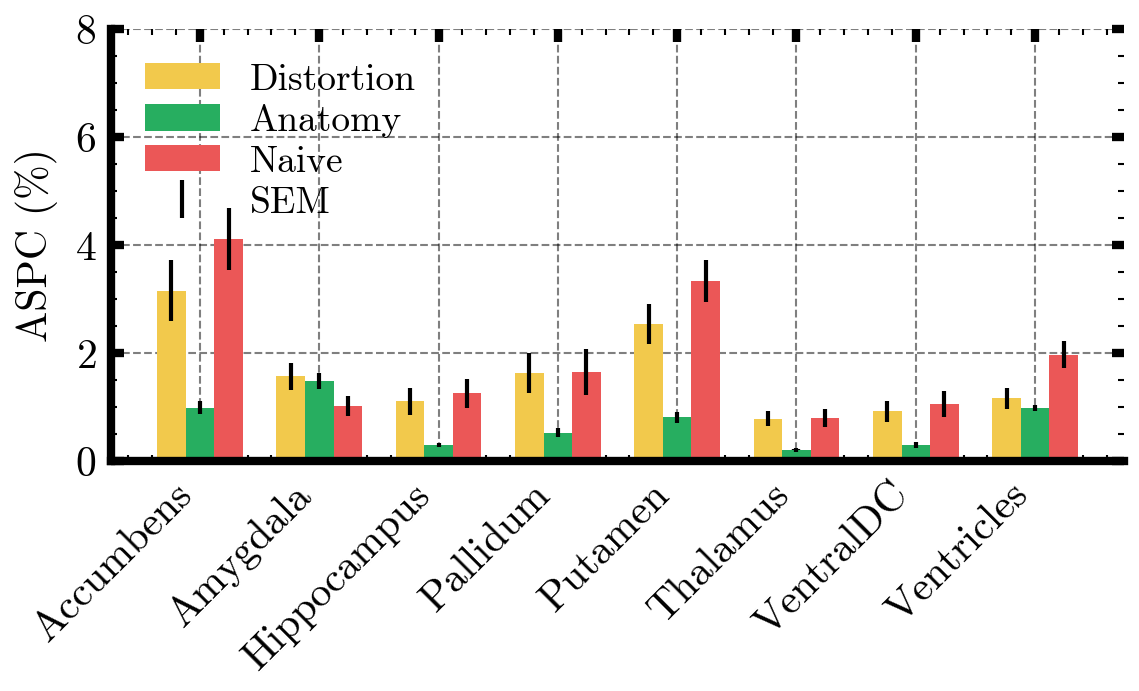}
    \caption{fBIRN (18 distortion-only pairs)}
    \label{fig:real_left}
\end{subfigure}
\hfill
\begin{subfigure}[t]{0.495\textwidth}
    \centering
    \includegraphics[width=\linewidth]{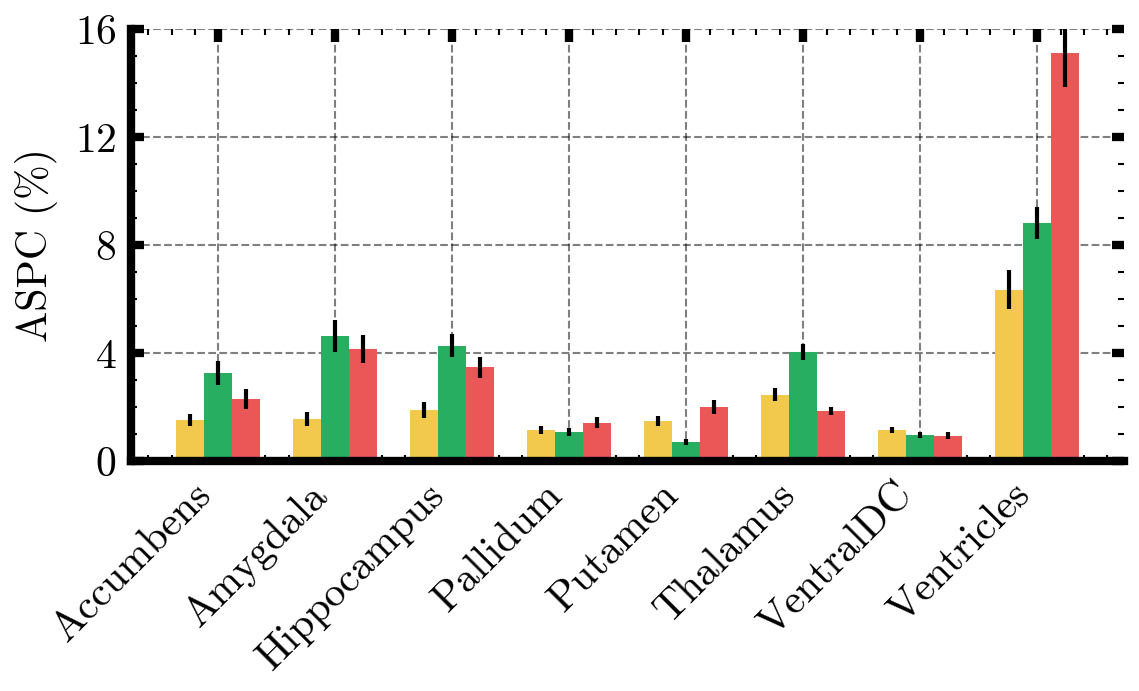}
    \caption{ADNI (40 longitudinal AD pairs)}
    \label{fig:real_right}
\end{subfigure}
\caption{Volume change (\texttt{ASPC}) in real data, captured by \texttt{Naive}, which includes distortion and atrophy; and separate $\texttt{DisMorph}_{\phi_d}$ and $\texttt{DisMorph}_{\phi_a}$ fields.
}
\label{fig:real_examples}
\end{figure}

\begin{figure}[t]
\centering
\includegraphics[width=\textwidth]{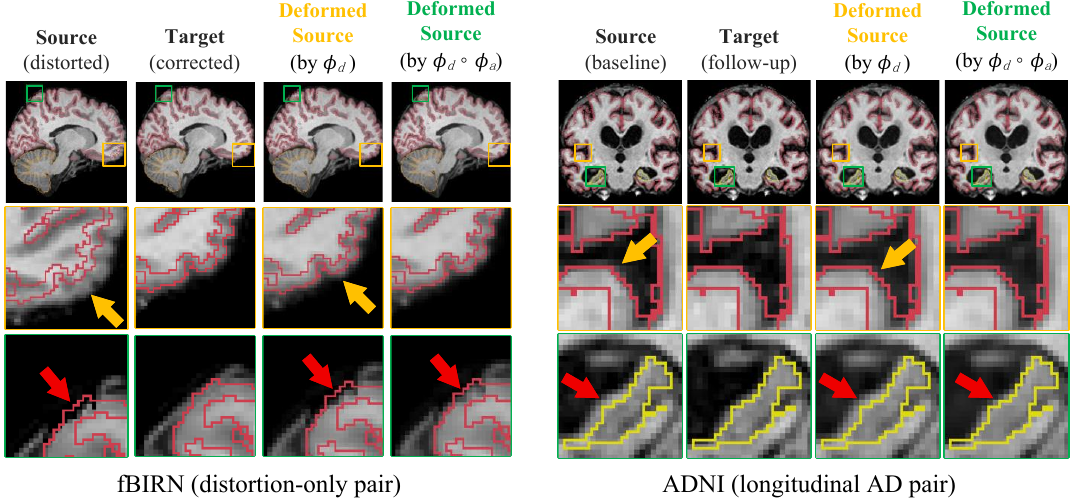}
\caption{Longitudinal registration examples. We overlay target-image label contours on all images (hippocampus in yellow; gray matter in red) and highlight changes with arrows. \textit{Left:} \method{} detects distortion. The example also shows a skull-stripping inconsistency between the two images, which \method{} interprets as apparent atrophy. \textit{Right:} \method{} captures subtle MRI-induced misalignment in the distortion field (row 2) while preserving hippocampal atrophy in the atrophy field (row 3).
} \label{fig:qualitative}
\end{figure}

\textbf{Data.} We evaluate {\method} on two real datasets: fBIRN~\cite{keator_function_2016} and ADNI~\cite{jack_jr_alzheimers_2008}.
First, we select one scan from each of 18 fBIRN subjects that we distortion-correct with \texttt{mri\_gradunwarp} using the manufacturer's coefficient files~\cite{jovicich2006reliability}. We pair up the original and gradient-corrected images; their geometric differences arise only from distortion (Fig.~\ref{fig:distortion}).

Second, we select longitudinal image pairs acquired two years apart from 40 ADNI 1 (1.5T) subjects diagnosed with AD. These images have been distortion-corrected in preprocessing. Although some residual, differential distortion typically remains, we expect the dominant geometric effect to be atrophy.

\noindent\textbf{Setup.} We compute per-structure \texttt{ASPC} between moving label maps and those moved by \texttt{Naive} registration $\phi$, the \method{} distortion field $\phi_d$ only, and the \method{} anatomical field $\phi_a$ after removing the distortion component.

\noindent\textbf{Results and Discussion.}
Fig.~\ref{fig:real_left} reports \texttt{ASPC} measured from the three deformation fields for fBIRN~\cite{keator_function_2016} and Fig.~\ref{fig:real_right} for ADNI~\cite{jack_jr_alzheimers_2008} across subcortical structures. For the fBIRN pairs, \method{} correctly attributes the majority of geometric differences to distortion. The residual signal in the atrophy field is minimal and might arise from preprocessing errors. We show an example in Fig.~\ref{fig:qualitative} (left), in which the distortion field (row 2) explains nearly all geometric differences, such that the source image deformed by the distortion field aligns closely with the target image, leaving an almost zero atrophy field. In contrast, row 3 highlights an inconsistency in skull-stripping, where tissue retained in the target image is absent from the source image. The model assigns this localized tissue loss to the atrophy field, likely because it resembles anatomical change rather than technical distortion.
For AD pairs, Fig.~\ref{fig:real_right} shows that the atrophy field captures the bulk subcortical volume change, consistent with the expected disease trajectories~\cite{tang_diffeomorphometry_2015,fu_synthesizing_2025}. The example in Fig.~\ref{fig:qualitative} (right) contains a large atrophy component and a small distortion component. The hippocampal volume loss visible after two years (row 3) remains preserved after applying the estimated distortion field, indicating that \method{} separates the smooth MRI-induced distortion from localized tissue change. Although trained solely on synthetic data, \method{} generalizes well to both real datasets and produces a biologically plausible decomposition of technical and anatomical deformation.

\section{Conclusion}
We introduce \method{}, the first registration framework that explicitly decomposes longitudinal deformation into technical and anatomical transforms. Across simulated and real datasets, our results demonstrate that this decomposition improves estimation of tissue change while isolating technical deformation. The ADNI analysis suggests that residual scanner-induced distortion persists despite standard vendor correction.
The present study evaluates the model's ability to isolate gradient nonlinearity. Future work will extend the evaluation to additional distortion sources, including B$_0$ field inhomogeneity and chemical shift~\cite{van_der_kouwe_brain_2008}. We also plan to investigate the impact of deformation decomposition on downstream morphometric analyses and statistical models of disease progression.
Explicitly simulating the physical origin of deformation in training may provide value to longitudinal morphometry in heterogeneous, retrospective, and multi-site imaging studies by complementing vendor corrections.

\begin{credits}
\subsubsection{\ackname} JF was supported by a WASP International Postdoctoral Scholarship
from the Knut and Alice Wallenberg Foundation. This work was supported
in part by NIH grants BICAN UM1 MH134812 and UM1 MH130981; BRAIN CONNECTS U01 NS132181 and UM1 NS132358; NIBIB R01 EB023281, R01 EB033773, R21 EB018907, R01 EB019956, and P41 EB030006; NICHD R00 HD101553; NIA R21 AG093037, R21 AG082082, R01 AG064027, R01 AG016495, and R01 AG070988; NIMH R01 MH123195, RF1 MH121885 and RF1 MH123195; NINDS U24 NS135561, R01 NS070963, R01 NS083534, R01 NS105820, and R25 NS125599; Blueprint for Neuroscience Research U01 MH093765; and Shared Instrumentation Grants S10 RR023401, S10 RR019307, and S10 RR023043. The project benefited from computational hardware provided by the Massachusetts Life Sciences Center.

\subsubsection{\discintname}
MH maintains a consulting relationship with Neuro42, Inc. BF is an advisor to DeepHealth. Their interests are reviewed and managed by Massachusetts General Hospital and Mass General Brigham in accordance with their conflict-of-interest policies. The authors have no other interests to declare that are
relevant to the content of this article.
\end{credits}
%
%
%
\clearpage 
\bibliographystyle{splncs04}
\bibliography{ref}
\end{document}